# Transient microscopy for measuring heat transfer in single cells

Pei Song[†1], He Gao[†1], Miao Zhang[1], Fan Yang[1], Shan-Shan Li[1], Bin Kang*[1], Jing-Juan Xu*[1] and Hong-Yuan Chen*[1]

[1]State Key Laboratory of Analytical Chemistry for Life Science and Collaborative Innovation Center of Chemistry for Life Sciences, School of Chemistry and Chemical Engineering, Nanjing University, 210023, China.

**Author Information**

[†]These authors contributed equally to this work.

*Correspondence and requests for materials should be addressed to B.K. (binkang@nju.edu.cn), J.-J.X. (xujj@nju.edu.cn) or H.-Y.C. (hychen@nju.edu.cn).

**Abstract**

Heat transfer and dissipation exists in almost any physical, chemical or biological systems. Cells, as the basic unit of life, undergo continuous heat transfer and dissipation during their metabolism. The heat transfer and dissipation within cells related to not only fundamental cellular functions and biochemical reactions, but also several important applications including heat-induced control of biological processes and treatment of diseases. Unfortunately, thus far, we still know very little about the heat transfer and dissipation properties at cellular or subcellular level. Here, we demonstrated a methodology of transient microscopy to map the heat transfer coefficients in single cells, with a temporal resolution of ~5 μs and a spatial resolution of ~250 nm (close to diffraction limit). The heat transfer coefficients of different location within single cells were obtained for the first time, the inner part of cells exhibited nonuniform heat transfer properties, and it suggested a self-consistent heat transfer regulation that responds to environmental temperature.

**Introduction**

Heat transfer and dissipation is critical in ensuring a stable physical[1], chemical[2], or biological system[3]. Warm-blooded mammals, such as humans, consume a large part of energy for heat[4,5] and maintain body temperature via balancing heat generation and dissipation[3,6]. Heat imbalance may cause an abnormal body temperature and even result in an irreversible physiological injury[6,7]. Cells, as the basic unit of life where most biochemical reactions occur, undergo continuous heat generation and dissipation[4,8]. The process of heat transfer and dissipation within the intracellular environment related on not only the fundamental cellular reactions,[4] metabolisms[5] and functions[9], but also several important applications including heat-induced control of gene expression[10], cancer metabolism[11] and selective treatment of disease[12,13]. Unfortunately, the heat transfer and dissipation properties at the cellular or subcellular level are nearly entirely unknown. Thus far, we still do not know the basic physical parameters of heat transfer within cellular environment, like heat transfer coefficient;

and we also do not know if cells could tune their heat dissipation capability under different environmental temperatures, just like human body does.

Recently, the local temperature distribution within cells has been tentatively explored via several promising techniques[14-17], with either high sensitivity of 1–2 mK[14] at nanoscale or high temporal resolution of several milliseconds[18]. Compared to measuring temperature, which is a state variable, the measurement of heat dissipation at nanoscale remains much more challenging[19], especially in a single cell. The main challenge is that measuring heat dissipation requires breaking the thermal equilibrium first, creating a heat transfer process at a subcellular level, and tracking the heat dissipation dynamics before the thermal equilibrium was rebuilt. During the measurement, the disturbance to the local thermal environment should be as low as possible to avoid the effects from temperature changes.

Here, we demonstrate a transient heat-transfer microscopy to map the heat transfer and dissipation properties within single cells by measuring the cooling time of a heated gold nanoparticle. The proposed method enabled visualizing heat dissipation within a single cell, with a temporal resolution of ~5 μs and a spatial resolution of ~250 nm (close to diffraction limit). The inner part of cells exhibited nonuniform heat transfer properties and a self-consistent heat transfer regulation that responds to environmental temperature. Based on these findings, we hypothesized that this cellular-level heat regulation existed in all homoeothermic animals and was an important part of body thermoregulation.

**Results**

**Principle of transient heat-transfer microscopy.** This transient heat-transfer microscopy is based on a time-correlated pump-probe principle and a wide-field imaging configuration (Fig. 1). A 532 nm laser with 5–7 ns pulses that match the maximum plasmon absorption band of gold nanoparticles, was used as a pump beam to heat the gold nanoparticle to a "hot" state, and then a ~5 μs-width pulsed white light was used as a probe beam to follow the cooling process. The sequences of pump and probe pulses were synchronized with a variable delay Δt. At time $t = 0$, the gold nanoparticle was rapidly heated by the pump beam to a "hot state" in several nanoseconds[20-22], and a localized temperature difference $\Delta T$ was generated, thereby causing a refractive index change $\Delta n$ in surrounding medium and formed a localized photothermal "nanolens"[23-25] that could be optically detected. The cooling process of the "hot" gold nanoparticle could be profiled by detecting the scatting signal of the probe beam at different delays $t = \Delta t$. Based on Newton's law of cooling, the thermal time constant $\tau$ and heat transfer coefficient $h_m$ of the surrounding medium could be expressed by the following equations[26, 27]:

$$\Phi(t) = A \cdot B \cdot C \frac{Q}{\rho_{Au} c_{Au} V_{Au}} e^{-\frac{t}{\tau}}, \tag{1}$$

$$\tau = \frac{\rho_{Au} c_{Au} V_{Au}}{h_m A_s}, \tag{2}$$

where $\Phi(t)$ is the detected optical signal at time $t$; $A$, $B$, and $C$ are constants; $Q$ is the pump energy; and $\rho_{Au}$, $c_{Au}$, $V_{Au}$, and $A_s$ are the density, specific heat capacity, body volume, and surface area of gold nanoparticles, respectively.

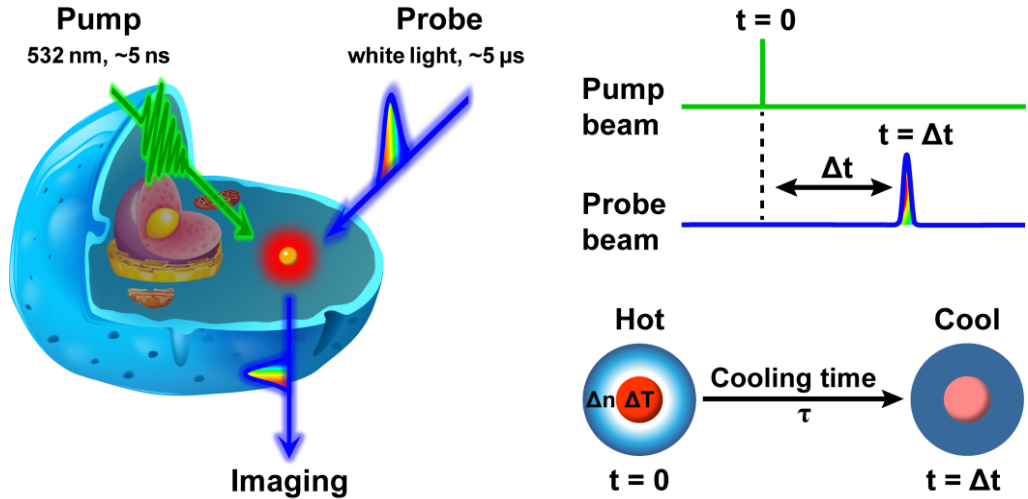

**Figure 1. Transient heat transfer microscopy.** A 532 nm pulsed laser (5–7 ns, 20 Hz) was used as the pump beam to heat a gold nanoparticle (yellow sphere with red glow) within a single cell at time $t = 0$. The cooling process was optically monitored by using a pulsed white light as the probe beam at variable $t = \Delta t$. The heat transfer coefficient ($h$) of the localized cellular medium could be obtained from the cooling time ($\tau$) according to Equation 2.

**Heat transfer coefficients in liquids.** We measured the heat transfer coefficient in a liquid medium with single gold nanoparticles to validate the principle experiment. The detected signal $\Phi(t)$ was defined as a relative change in plasmonic scattering intensity: $\Phi(t) = (I_{\Delta t} - I_s)/I_s$, which originated from the refractive index change $\Delta n$ caused by temperature change $\Delta T$, where $I_{\Delta t}$ is the scattering intensity at $t = \Delta t$, and $I_s$ is the steady-state scattering intensity without a pump. Heat transfer in air was undetectable in the current system (Fig. 2a) because the $\Phi(t)$ generated in air was close to the detection limit (~±0.02) even with a strong pump. The $\Phi(t)$ was undetected in the water medium when the pump beam was switched off because no "nanolens" effect was generated from surrounding nanoparticles without heating (Fig. 2b). The cooling process of a single gold nanoparticle in water and glycerol was successfully tracked when the pump and probe beams were switched on (Fig. 2c). The time profile indicated that the heat transfer coefficients for water and glycerol were 469±24 and 226±13 W·m$^{-2}$·K$^{-1}$, respectively. The influence of pump energy $Q$ on $\Phi(t)$ was also evaluated. The initial $\Phi(0)$ at $\Deltaت = 0$ exhibited a linear correlation with the pump energy (Fig. 2d), which matched well with our theory. The dependence of the measured $h$ on the measurement conditions, including the pump energy, physical size, and surface chemistry of nanoparticles, were carefully investigated (Figs. 2e–g). The results showed that the final $h$ did not depend on pump energy, particle size, or surface conjugation. These results were reasonable because heat transfer coefficient $h$ is an instinct parameter of the medium that should not depend on measurement conditions. Fig. 2e demonstrates that an increase in pump energy induced a slight increase in $h$, especially at relatively high pump energy of 1.2 mJ. This condition could be attributed to the local temperature change in the surrounding medium under a high pump energy because $h$ is a temperature-dependent parameter for most liquids[26]. In subsequent experiments, pump energy (<0.5 mJ) and temperature change on the sample (<0.1 K) were carefully controlled to minimize thermal perturbation. Under these conditions, the localized temperature change around gold nanoparticles was <5 K, and the detection limit of our method was ~0.5 K in water and ~0.2 K in glycerol.

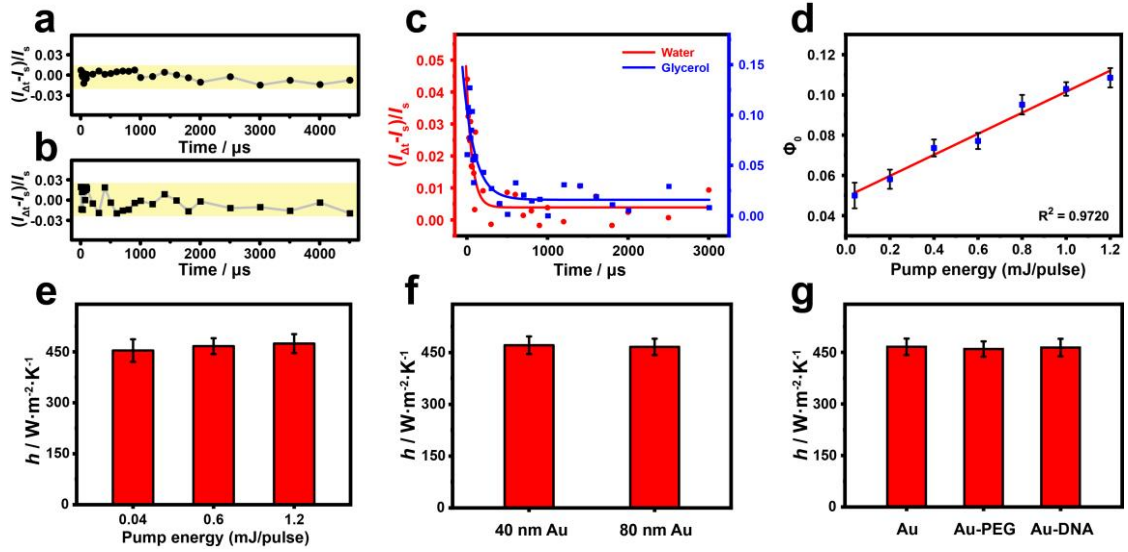

**Figure 2. Single-particle-based heat transfer measurement. a,b**, Curves of $\Phi(t)$ changes with time in air with pump laser on (**a**) and in water with pump laser off (**b**). **c**, Cooling time curves of gold nanoparticle in the surrounding medium of water (red curve) and glycerol (blue curve). **d**, Initial $\Phi(0)$ at $t = 0$ versus the pump energy. **e-g**, Heat transfer coefficients of water was measured under different pump energies (**e**), particle sizes (**f**), and surface conjugations (**g**).

**Heat transfer coefficients in single cells.** We demonstrated that our method could be used to map heat transfer in cells by using HeLa cell line as model. The control cells without gold nanoparticles exhibited very low background signals ($\Phi$ = ~0.02–0.05) closed to the detection limit (±0.02) given their low absorption cross section at 532 nm, whereas the cells with gold nanoparticles showed strong signals. In Fig. 3, images of $\Phi(t)$ at different delay times could be obtained by applying $\Phi(t) = (I_t - I_s)/I_s$ to every pixel of the dark-field (DF) image of cells with gold nanoparticles (Figs. 3a and b). The images of $\Phi(t)$ present the decay correlated with the cooling process of each nanoparticle. The time profile of each pixel in the images of the $\Phi(t)$ could be exponentially fitted similar to the typical cases illustrated (Fig. 3c). The value of $h$ at every position of the $\Phi(t)$ images could be calculated and remapped as an $h$ map (Fig. 3d). The $h$ map showed that cells exhibited nonuniform heat transfer properties. The heat transfer coefficient $h$ covered a wide range from 129 W·m$^{-2}$·K$^{-1}$ to 334 W·m$^{-2}$·K$^{-1}$, with an average $h$ = 241 W·m$^{-2}$·K$^{-1}$. Unlike in solution, the inner part of cells was a sticky and crowded environment[28, 29], which contained numerous types of macromolecules and subcellular organelles[30]. This heterogeneous physicochemical environment might lead to the observed heterogeneity on heat transfer properties.

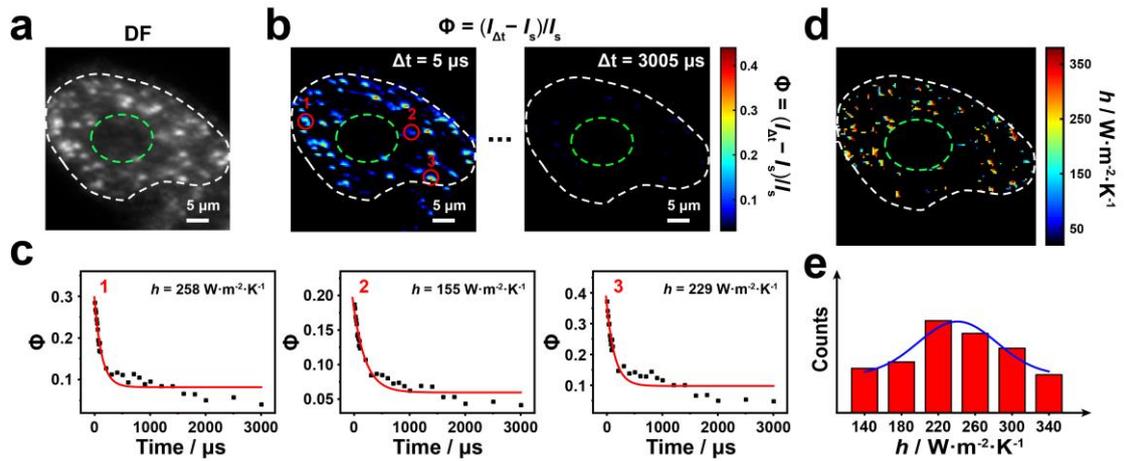

**Figure 3. Map heat transfer in single cells. a**, DF image of a HeLa cell at time Δt = 5 μs. **b**, $\Phi(t)$ snapshot of the HeLa cell at Δt = 5 μs and Δt = 3005 μs during the cooling process. **c,** Cooling curves of three typical regions in the HeLa cell. **d,e,** Image (**d**) and corresponding histogram (**e**) of heat transfer coefficient distribution in the HeLa cell. The white dash lines in **a**, **b**, and **d** indicate cell contour, and the green dash lines indicate cell nuclei.

**Heat regulation at cellular level.** We explored heat transfer in cells under different environmental temperatures (Fig. 4). From a room temperature of 25 ℃ to a human body temperature of 37 ℃, the average heat transfer coefficient of cells showed a quasi-linear increase from 231±5 W·m$^{-2}$·K$^{-1}$ to 266±6 W·m$^{-2}$·K$^{-1}$, with *dh/dT* = 2.9 W·m$^{-2}$·K$^{-2}$ (Fig. 4a). This phenomenon was likely a regular temperature correlation, as reflected in a cell culture medium with *dh/dT* = 3.3 W·m$^{-2}$·K$^{-2}$ (Fig. 4b). When the environmental temperature increased from 37 ℃ to 42 ℃, the average heat transfer coefficient of cells remarkably increased to 313±8 W·m$^{-2}$·K$^{-1}$, with roughly estimated *dh/dT* = 9.5 W·m$^{-2}$·K$^{-2}$ (Fig. 4a). The heat dissipation histograms (Figs. 4c and d) clearly denote the increase in heat transfer coefficient within cells from 37 ℃ to 42 ℃. Most cell cultures required a temperature of approximately 37 ℃, and a long-term exposure to a high temperature of 42 ℃ would result in cell injury and even cell death[31]. In our measurement, the duration of high temperature was carefully controlled to several minutes to avoid irreversible cell injuries. Cell morphology remained unchanged during measurement, and consistent results could be observed when cells were cooled down to the normal temperature. Cells presented a self-consistent regulation on their heat transfer properties responsive to environmental temperature. This unique property could be conducive to maintaining the intracellular temperature within a reasonable range, although the mechanisms were unclear. The current work focused on developing fundamental methodologies and did not involve large-scale cell screening. On the basis of these results, it is reasonable to hypothesize that this cellular-level heat regulation might exist in all homoeothermic animals.

Further systematic verifications on a large range of cell lines are required to test this hypothesis.

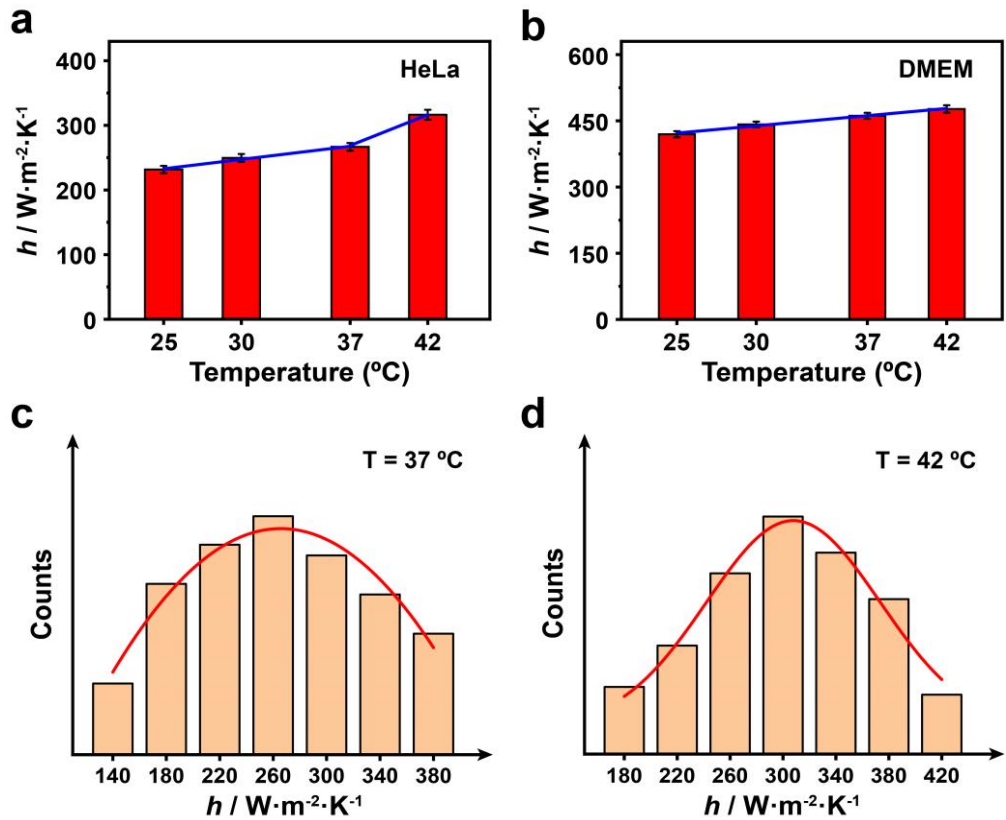

**Figure 4. Temperature-dependent heat transfer in single cell**. **a**, Average heat transfer coefficient of HeLa cell at environmental temperatures of 25, 30, 37, and 42 ℃. **b**, Heat transfer coefficients of HeLa cell culture medium at environmental temperatures of 25, 30, 37, and 42 ℃. **c–d**, corresponding histograms of heat transfer coefficient distribution in the HeLa cell at 37 and 42 ℃.

**Discussion**

The present measurement was based on single gold nanoparticles, with 40 or 80 nm in diameter. Spatial resolution mainly depends on the thickness of the thermal diffusion layer in surrounding medium. In our system, the thickness of the thermal layer was related to the power of pump laser and heat diffusion time. The scale of the thermal field could be decreased to <100 nm by optimizing the pump power. The heat dissipation processes tracked in this work were typically with a thermal time constant <1 ms, and the temporal resolution (i.e., time gate width) of our transient microscopy was ~5 μs. Compared with previous techniques that measure steady-state temperatures, our strategy reached the highest temporal resolution to date. The combination of diffraction-limit imaging and transient temporal resolution capabilities in our method offered unique advantages to map the ultrafast dynamic biochemical events at cellular or subcellular level, with an nm-μs spatial-temporal resolution simultaneously. Heat dissipation within subcellular organelles could thus be explored by conjugating particles with specific targeting ligands[32, 33]. The combination of the proposed technique and precision temperature sensing[14, 34] might further enable a comprehensive evaluation of thermogenesis and dissipation at the single-cell level.

The present observations generated several unknown questions. First, extensive validations are required to verify whether this cellular-level heat regulation exists in homoeothermic and poikilothermic animals. Second, the molecular mechanism and pathways of cellular heat regulation are unknown. Fully unfolding the answers definitely requires efforts of biologists to yield relevant molecular "sensors" and "actuators" and determine the pathways in changing heat transfer. Third, heat dissipation at cell level may relate to diseases, such as cancers, and behavior under extreme conditions, such as Olympic athletes.

# References

1. Liao, A. et al. Thermal dissipation and variability in electrical breakdown of carbon nanotube devices. *Phys. Rev. B* **82**, 205406 (2010).
2. Meyer, J. & Reuter, K. Modeling heat dissipation at the nanoscale: An embedding approach for chemical reaction dynamics on metal surfaces. *Angew. Chem. Int. Ed.* **53**, 4721-4724 (2014).
3. Niedermann, R. et al. Prediction of human core body temperature using non-invasive measurement methods. *Int. J. Biometeorol.* **58**, 7-15 (2014).
4. Himms-Hagen, J. Cellular thermogenesis. *Annu. Rev. Physiol.* **38**, 315-351 (1976).
5. Ricquier, D. Fundamental mechanisms of thermogenesis. *C. R. Biol* **329**, 578-586 (2006).
6. Houdas, Y. & Ring, E. Human body temperature: Its measurement and regulation. (Springer Science & Business Media, 2013).
7. Bowler, K., Laudien, H. & Laudien, I. Cellular heat injury. *J. Therm. Bio.* **8**, 426-430 (1983).
8. James, A.M. Thermal and energetic studies of cellular biological systems. (Butterworth-Heinemann, 2016).
9. Wegner, N.C., Snodgrass, O.E., Dewar, H. & Hyde, J.R. Whole-body endothermy in a mesopelagic fish, the opah, Lampris guttatus. *Science* **348**, 786 (2015).
10. Kamei, Y. et al. Infrared laser–mediated gene induction in targeted single cells in vivo. *Nat. Methods* **6**, 79 (2008).
11. Vreugdenburg, T.D., Willis, C.D., Mundy, L. & Hiller, J.E. A systematic review of elastography, electrical impedance scanning, and digital infrared thermography for breast cancer screening and diagnosis. *Breast Cancer Res. Tr.* **137**, 665-676 (2013).
12. Schroeder, A. et al. Treating metastatic cancer with nanotechnology. *Nat. Rev. Cancer* **12**, 39 (2011).
13. Dutz, S. & Hergt, R. Magnetic nanoparticle heating and heat transfer on a microscale: Basic principles, realities and physical limitations of hyperthermia for tumour therapy. *Int. J. Hyperther.* **29**, 790-800 (2013).
14. Kucsko, G. et al. Nanometre-scale thermometry in a living cell. *Nature* **500**, 54-58 (2013).
15. Okabe, K. et al. Intracellular temperature mapping with a fluorescent polymeric thermometer and fluorescence lifetime imaging microscopy. *Nat. Commun.* **3**, 705 (2012).
16. Yang, J.-M., Yang, H. & Lin, L. Quantum dot nano thermometers reveal heterogeneous local thermogenesis in living cells. *ACS Nano* **5**, 5067-5071 (2011).
17. Nakano, M. & Nagai, T. Thermometers for monitoring cellular temperature. *J. Photoch. Photobio. C* **30**, 2-9 (2017).
18. Chen, X. et al. Imaging the transient heat generation of individual nanostructures with a mechanoresponsive polymer. *Nat. Commun.* **8**, 1498 (2017).
19. Hoogeboom-Pot, K.M. et al. A new regime of nanoscale thermal transport: Collective diffusion increases dissipation efficiency. *Proc. Natl. Acad. Sci. U.S.A.* **112**, 4846-4851 (2015).
20. Qin, Z. & Bischof, J.C. Thermophysical and biological responses of gold nanoparticle laser heating. *Chem. Soc. Rev.* **41**, 1191-1217 (2012).
21. Rashidi-Huyeh, M. & Palpant, B. Thermal response of nanocomposite materials under pulsed laser excitation. *J. Appl. Phys.* **96**, 4475-4482 (2004).
22. Hodak, J.H., Henglein, A. & Hartland, G.V. Photophysics of nanometer sized metal particles: Electron−phonon coupling and coherent excitation of breathing vibrational modes. *J. Phys. Chem.*

*B* **104**, 9954-9965 (2000).

23. Chen, Z. et al. Imaging local heating and thermal diffusion of nanomaterials with plasmonic thermal microscopy. *ACS Nano* **9**, 11574-11581 (2015).
24. Heber, A., Selmke, M. & Cichos, F. Thermal diffusivity measured using a single plasmonic nanoparticle. *Phys. Chem. Chem. Phys.* **17**, 20868-20872 (2015).
25. Selmke, M., Braun, M. & Cichos, F. Photothermal single-particle microscopy: Detection of a nanolens. *ACS Nano* **6**, 2741-2749 (2012).
26. Bergman, T.L. & Incropera, F.P. Fundamentals of heat and mass transfer. (John Wiley & Sons, 2011).
27. Vollmer, M. Newton's law of cooling revisited. *Europ. J. Phys.* **30**, 1063 (2009).
28. Kuimova, M.K. et al. Imaging intracellular viscosity of a single cell during photoinduced cell death. *Nat. Chem.* **1**, 69-73 (2009).
29. Peng, X. et al. Fluorescence ratiometry and fluorescence lifetime imaging: Using a single molecular sensor for dual mode imaging of cellular viscosity. *J. Am. Chem. Soc.* **133**, 6626-6635 (2011).
30. Karp, G. Cell and molecular biology: Concepts and experiments 4th edition with plus set. (John Wiley & Son, 2006).
31. Lepock, J.R. Cellular effects of hyperthermia: Relevance to the minimum dose for thermal damage. *Int. J. Hyperther.* **19**, 252-266 (2003).
32. Kang, B., Mackey, M.A. & El-Sayed, M.A. Nuclear targeting of gold nanoparticles in cancer cells induces DNA damage, causing cytokinesis arrest and apoptosis. *J. Am. Chem. Soc.* **132**, 1517-1519 (2010).
33. Ma, X., Gong, N., Zhong, L., Sun, J. & Liang, X.-J. Future of nanotherapeutics: Targeting the cellular sub-organelles. *Biomaterials* **97**, 10-21 (2016).
34. Toyli, D.M., de las Casas, C.F., Christle, D.J., Dobrovitski, V.V. & Awschalom, D.D. Fluorescence thermometry enhanced by the quantum coherence of single spins in diamond. *Proc. Natl. Acad. Sci. U.S.A.* **110**, 8417-8421 (2013).

## Acknowledgements

This work was mainly supported by the National Natural Science Foundation of China (21327902, 21535003, and 21675081), State Key Laboratory of Analytical Chemistry for Life Science (5431ZZXM1715), and Priority Academic Program Development of Jiangsu Higher Education Institutions.

## Author Contributions

P.S. and H.G contributed equally to this work. H.-Y.C. J.-J.X. and B.K. conceived the research.